# Experiments on Manual Thesaurus based Query Expansion for Ad-hoc Monolingual Gujarati Information Retrieval Tasks

Hardik Joshi, Dr. Jyoti Pareek

*Abstract*-- **In this paper, we present the experimental work done on Query Expansion (QE) for retrieval tasks of Gujarati text documents. In information retrieval, it is very difficult to estimate the exact user need, query expansion adds terms to the original query, which provides more information about the user need. There are various approaches to query expansion. In our work, manual thesaurus based query expansion was performed to evaluate the performance of widely used information retrieval models for Gujarati text documents. Results show that query expansion improves the recall of text documents.**

## I. INTRODUCTION

Information retrieval (IR) is concerned with representing, searching, and manipulating large collections of electronic text data generally in unstructured form. IR is the discipline that deals with retrieval of unstructured data or partially structured data, especially textual documents, in response to a set of query or topic statement(s), which may itself be unstructured. [6] The typical interaction between a user and an IR system can be modeled as the user submitting a query to the system; the system returns a ranked list of relevant documents, with the most relevant at top of the list.

It is required to develop effective methods of automated IR due to the tremendous explosion in the amount of text documents and growing number of document sources on the Internet. Over the last few years, there has been a significant growth in the amount of text documents in Indian languages. Researchers have been performing IR tasks in English and European languages since many years [2], [15], efforts are being made to encourage IR tasks for the Indian Languages[5], [14].

Most of the IR research community uses resources known as test collection [12] . Since 1990's, TREC [15] is conducting evaluation exercises using test collections. The classic components of a test collection are :

   a) A collection of documents; each document is identified uniquely by docid.
   b) A set of queries (also referred as topics); each query is uniquely identified by a qid.
   c) A set of relevance judgements (also referred as qrels) that consists of a list of (qid,docid) pairs detailing the relevance of documents to topics.

In this paper, we have described the ad hoc monolingual IR task performed over Gujarati language collection. In ad hoc querying, the user formulates any number of arbitrary queries but applies them to a fixed collection. [7] We have considered Gujarati language, the reason being no such tasks have been performed for Gujarati language, although some work is carried out for Bengali, Hindi and Marathi languages.[10], [13] Apart from this Gujarati is spoken by nearly 50 Million people over the world and is an official language for the state of Gujarat. [9]

The rest of the paper is organized as follows: Section 2 describes briefly query expansion and various techniques applied for the same. Section 3 gives an overview of the experimental setup used to perform ad hoc task. Section 4 presents an overview of the evaluation conducted. Section 5 presents the results obtained during the experiment and finally Section 6 concludes the experiment.

## II. OVERVIEW OF QUERY EXPANSION

The information retrieval system can lead to inaccurate results due to the incorrect query formed by a few keywords. The actual information need of any user can be difficult to model with few keywords in the query. The most common ambiguity occurs due to term mismatch wherein the indexer may use a different set of words while the query may contain synonyms of the indexed words. For example words like 'tv' and 'television' represent the same document but the indexer might have used 'television' while the query may contain words like 'tv' resulting in decrease of recall. In interactive search engines, users often attempt to address such problems by refining the query themselves. Query refinement can be made fully automatically or with the inputs from user.

Such problems can be tackled either using global methods or local methods. Global methods are techniques for expanding or reformulating the query terms independent of the query and the results returned from it whereas the local methods adjust a query relative to the document that initially appear to match the query. The followings are various techniques for global methods and local methods for query expansion [11]:

Global Methods:
   a) QE using thesaurus or WordNet
   b) QE using automatic thesaurus Generation
   c) QE with Spelling correction

Local Methods:
   a) QE with Relevance Feedback
   b) QE with Pseudo Relevance Feedback
   c) QE with Indirect Relevance Feedback

We have applied thesaurus based query expansion to Gujarati queries. On an average 6 terms were added to reformulate each query. Synonyms were looked up from the Gujarati Lexicon and were added manually in the queries.

## III. EXPERIMENTAL SETUP

### 1) Overview of the Corpus

The test collection used for this experiment is the collection made available during the FIRE 2011. [5] The details of Gujarati Collection are mentioned in Table I. The collection was created from the archives of the daily newspaper, "Gujarat Samachar" from 2001 to 2010. Each document represents a news article from "Gujarat Samachar". The average tokens per document are 445.

TABLE I
STATISTICS OF THE GUJARATI CORPUS

| Particulars | Quantity |
| --- | --- |
| Size of Collection | 2.7 GB |
| Number of text Documents | 3,13,163 |
| Size of Vocabulary | 20,92,619 |
| Number of Tokens | 13,92,72,906 |

The corpus is coded in UTF-8 and each article is marked up using the following tags:

<DOC> : Starting of the document
<DOCNO> </DOCNO> : Unique identifier of the document
<TEXT> </TEXT> : Contains the document text
</DOC> : Ending tag of the document

### 2) Queries

The IR models were tested against 50 different queries in Gujarati language. Following the TREC model [16], each query is divided into three sections: the title (T) which indicates the brief title, the description (D) which gives a one-sentence description and the narrative part (N), which specifies the relevance assessment criteria. Below is an example of a single query in the collection of 50 queries.
<top>
<num> </num> : Unique identifier of the Query
<title> </title> : Title of the Query
<desc> </desc> : Short description of the Query
<narr> < narr> : Detailed Query in narrative form
</top>

### 3) IR Models

In the experiment, we have compared various models that are mostly used in test collection evaluation exercise. We considered classical models like Term Frequency Inverse Document Frequency (TF_IDF), language models like Hiemstra Language Model (Hiemstra_LM) [8], probabilistic models like Okapi (BM25) [1], Divergence from Randomness (DFR) group of models like Bose-Einstein model for randomness which considers the ratio of two Bernoulli's processes for first Normalization, and Normalization 2 for term frequency normalization (BB2), The DLH hyper-geometric DFR model (DLH) and its improvement (DLH13), Divergence From Independence model (DFI0), A different hyper-geometric DFR model using Popper's normalization (DPH) which is parameter free, DFR based hyper-geometric models which takes an average of two information measures (DFRee), Inverse Term Frequency model with Bernoulli after-effect (IFB2), Inverse Expected Document Frequency model with Bernoulli after-effect (In_expC2), Inverse Document Frequency model with Laplace after-effect (InL2), Poisson model with Laplace after-effect (PL2), a log-logistic DFR model ( LGD) [4] and an Unsupervised DFR model that computed the inner product of Pearson's $X^2$. The experiment used 21 different models to perform information retrieval of Gujarati text documents. Few of the models required parametric values; we have used the default values that are generally applied to similar tests.

## IV. EVALUATION

In earlier years, the IR systems were evaluated using measures like Precision, Recall and Fallout [12], where precision measures the fraction of retrieved documents that are relevant whereas recall measures the fraction of relevant documents retrieved and fallout measures the fraction of non-relevant documents retrieved. In recent years Mean Average Precision (MAP) values are considered to give the best judgment in the presence of multiple queries [11].

In our experiments, to evaluate the retrieval performance, we have used the recall values. We evaluated the results separately for title (T), combination of title and description (TD) and the combination of title, description and narration (TDN). Although MAP values are widely used in IR communities and has good discrimination and stability, it has been observed that Query Expansion techniques results in lower MAP values but it does improve the recall. Recall is the fraction of the documents that are relevant to the query that are successfully retrieved.

$$recall = \frac{|\{\text{relevant documents}\} \cap \{\text{retrieved documents}\}|}{|\{\text{relevant documents}\}|}$$

In our experiments, we have tested various models for recall values. The results obtained from tests performed on TD are more satisfying when compared to T or TDN.

## V. RESULTS

It has been observed that the combination of Title (T) and Description (D) gives better MAP values instead of using them separately. A baseline of MAP values was obtained from Gujarati IR tasks of various IR models [17]. MAP values obtained by performing query expansion fall below the baseline however we observed a significant increase in recall values for most of the models. A summary of results in terms of recall values obtained before and after query expansion are listed in table I.

| IR MODELS | Relevant | TD(Before Query Expansion) | | TD(After Query Expansion) | | Results |
|---|---|---|---|---|---|---|
| | | Relevant Retrieved | Average Retrieval (Percentage) | Relevant Retrieved | Average Retrieval (Percentage) | |
| BB2 | 1659 | 1195 | 72 | 1202 | 72.5 | Improvement |
| BM25 | 1659 | 1169 | 70.5 | 1176 | 70.9 | Improvement |
| DFI0 | 1659 | 1092 | 65.8 | 1107 | 66.7 | Improvement |
| DFRBM25 | 1659 | 1172 | 70.6 | 1178 | 71 | Improvement |
| DFRee | 1659 | 1058 | 63.8 | 1066 | 64.3 | Improvement |
| DirichletLM | 1659 | 1138 | 68.6 | 1143 | 68.9 | Improvement |
| DLH | 1659 | 1130 | 68.1 | 1145 | 69 | Improvement |
| DLH13 | 1659 | 1104 | 66.5 | 1108 | 66.8 | Improvement |
| DPH | 1659 | 1109 | 66.8 | 1143 | 68.9 | Improvement |
| IFB2 | 1659 | 1236 | 74.5 | 1244 | 75 | Improvement |
| In_expB2 | 1659 | 1218 | 73.4 | 1225 | 73.8 | Improvement |
| In_expC2 | 1659 | 1226 | 73.9 | 1231 | 74.2 | Improvement |
| LemurTF_IDF | 1659 | 1250 | 75.3 | 1252 | 75.5 | Improvement |
| PL2 | 1659 | 1150 | 69.3 | 1159 | 69.9 | Improvement |
| XSqrA_M | 1659 | 1070 | 64.5 | 1084 | 65.3 | Improvement |
| TF_IDF | 1659 | 1152 | 69.4 | 1164 | 70.2 | Improvement |
| Hiemstra_LM | 1659 | 1164 | 70.2 | 1158 | 69.8 | Fail |
| InB2 | 1659 | 1195 | 72 | 1176 | 70.9 | Fail |
| InL2 | 1659 | 1159 | 69.9 | 1151 | 69.4 | Fail |
| Js_KLs | 1659 | 1076 | 64.9 | 1076 | 64.9 | Fail |
| LGD | 1659 | 1089 | 65.6 | 1089 | 65.6 | Fail |

19

## VI. CONCLUSION

The investigations of our experiments show that query expansion does improve the recall values for most of the models except Hiermstra's manguage model, InB2, InL2, Js_KLs and LGD models.

## VII. ACKNOWLEDGEMENTS

This work was supported by the University Grants Commission (UGC) Minor Research Project, grant number: [F. No: 41-1360/2012 (SR)]. We would like to extend our gratitude to the FIRE forum for providing the researchers with the data/corpus made available free of cost and for providing the relevance judgments.